# Direct Manipulation of quantum entanglement from the non-Hermitian nature of light-matter interaction

Kangkang Li[1,2], Yin Cai[1,‡], Jin Yan[1], Zhou Feng[1], Fu Liu[1], Lei Zhang[1], Feng Li[1,†], Yanpeng Zhang[1,*]

[1]Key Laboratory for Physical Electronics and Devices of the Ministry of Education & Shaanxi Key Lab of Information Photonic Technique, Xi'an Jiaotong University, Xi'an 710049, China

[2]State Key Laboratory for Artificial Microstructure and Mesoscopic Physics and Frontiers Science Center for Nano-optoelectronics, School of Physics, Peking University, 100871 Beijing, China

Emails: ‡caiyin@xjtu.edu.cn; †felix831204@xjtu.edu.cn; *ypzhang@mail.xjtu.edu.cn.

**Biphoton process is an essential benchmark for quantum information science and technologies, while great efforts have been made to improve the coherence of the system for better quantum correlations. Nevertheless, we find that the non-Hermitian features induced by the atomic quantum interference could be well employed for the direct control of entanglement. We report the demonstration of exceptional point (EP) in biphotons by measuring the light-atom interaction as a natural non-Hermitian system, in which the electromagnetically induced transparency regime provides a powerful mechanism to precisely tune the non-Hermitian coupling strength. Such biphoton correlation is tuned within an unprecedented large range from Rabi oscillation to antibunching-exponential-decay, also indicating high-dimensional entanglement within the strong and weak light-matter coupling regimes. The EP at the transition point between the two regimes is clearly observed with the biphoton quantum correlation measurements, exhibiting a single exponential decay and manifesting the coalesced single eigenstate. Our results provide a unique method to realize the controllability of natural non-Hermitian processes without the assistance of artificial photonic structures, and paves the way for quantum control by manipulating the non-Hermitian features of the light-matter interaction.**

Non-Hermitian Hamiltonians, characterized by particle creation and decay via the interaction with external environment, are of great significance for the proper description of realistic physical systems[1-4]. Recently, a remarkable merit of the non-Hermitian systems is investigated at the exceptional point (EP), i.e., the non-Hermitian degeneracy at which both the real and imaginary parts of the system coalesce, displaying a single eigenstate[5-8]. The investigations of EP have resulted in interesting physics and applications such as enhanced sensing[9,10], loss-mediated lasing [11], laser mode selectivity[12], non-reciprocity[13], topological energy transfer[14], skin effect[15], divergent quantum metric[16], etc. In photonic systems, the EP was usually demonstrated and investigated in artificial structures such as microcavities[10,16,17] and waveguides[18], especially in those displaying parity-time (PT) symmetry which provide a straight forward way to precisely tune the coupling strength, the gain, or the dissipation. The EP is the phase transition point between the so-called exact (strong coupling) and broken (weak coupling) PT-symmetric states, which are characterized by a bifurcation of the real and imaginary parts of eigenstates, respectively.

On the other hand, it is well known that photonic systems are naturally non-Hermitian due to radiative decay and structural losses, which become particularly interesting in the two-level systems with light-matter interaction. Indeed, the strong coupling regimes have been demonstrated in atomic systems[19-21], exciton-polaritons[22-24], surface plasmon polaritons[25,26], etc. In principle, such systems are naturally-existing candidates for EP studies, as the EP could be characterized by precisely tuning the systems from strong to weak light-matter coupling regimes, requiring no artificial photonic structures. To demonstrate the EP inherently endowed within the natural light-matter coupling process, we need to include such natural processes, such as spontaneous radiation and absorption, as the coupling terms (off-diagonal elements) of the non-Hermitian Hamiltonian and to precisely tune the system coupling around the EP. Nevertheless, due to the difficulty of fine tuning the coupling strength and lifetime of the two-level systems, it seems impractical either to precisely reach the EP or to adjust the systems continuously around the EP. Therefore, even though the light-matter coupling process are naturally non-Hermitian, its non-Hermitian property is mostly employed only as an experimental platform for EP studies in artificially designed photonic structures[27,28]. Although the artificial structures exhibit EP, the experimental system works deterministically either in the strong or weak light-matter coupling regime, leaving the EP of the light-matter coupling itself not yet achieved.

Moreover, biphotons processes are essential for generating quantum entanglement and realizing quantum information tasks. Recently, nonlinearly multi-wave mixing processes in the natural non-Hermitian system with neutral atoms provide an efficient quantum platform to generate narrow bandwidth biphotons[29,30]. Generally, the interaction with environment introduces decoherence and gives rise to the destruction for quantum correlations[31]. Nevertheless, for dissipative atomic structures, the electromagnetic induced transparency (EIT), being the destructive interference between two quantum pathways[32-37], can be employed to fine tune the light-matter coupling between the atomic states and external optical fields, which allows actively controlling the produced quantum correlations. However, limited by the difficulty from co-existing linear and nonlinear responses, it remains a major challenge to realize fine tuning the coupling strength and lifetime to reach the EP within natural non-Hermitian systems, especially in nonlinear and entanglement generation processes.

In this work, we demonstrate the EP of natural light-matter interaction in the EIT-based spontaneous four-wave mixing (SFWM), which is characterized by time-energy entangled photon pairs. By precisely tuning the coupling strength with a control beam, the biphoton correlation changes from the strong coupling regime characterized by Rabi oscillation, to the weak coupling regime characterized by antibunching-exponential-decay. At the phase transition of the EP, a clear feature of single exponential decay is observed, manifesting a coalesced single eigenstate. The observations agree well with the theoretical model described by the non-Hermitian Hamiltonian. Our results unambiguously show that the non-Hermitian features of natural light-matter coupling systems can be well controlled by quantum optical processes, and such naturally-existing non-Hermitian systems can serve as novel platforms for modulating high-dimensional quantum information processing benefiting from the merits of the EPs.

In our scheme, with pump beam $E_1$ and coupling beam $E_2$ as shown in Fig. 1a, b, the time-energy entangled biphoton $E_S$ and $E_{AS}$ are generated via an EIT-based SFWM process within a typical four-level rubidium atomic configuration. The details of experimental implementation are described in the **Methods**. Physically, the interaction Hamiltonian of the biphoton processing is defined as $H_I = i\hbar\kappa\Phi\hat{a}_S^\dagger\hat{a}_{AS}^\dagger + H.C.$ with annihilation operators $\hat{a}_S^\dagger$ and $\hat{a}_{AS}^\dagger$ of $E_S$ and $E_{AS}$; $\kappa$ is nonlinear coupling coefficient and $\Phi$ is longitudinal detuning function; $H.C.$ is Hamiltonian conjugation. Since the waveform of biphoton correlation function $G^{(2)}$ is defined by the convolution of $\kappa$ and $\Phi$ (details in the Supplemental Material), the Rabi oscillation and group delay regimes of biphoton correlation are demonstrated in the strong coupling condition[35-39]. However, for observing the EP, the system need be characterized by precisely tuning coupling regime in a wide range, from strong coupling to EP, and to weak coupling regimes. In particular, the typical challenge is the nonlinear coupling coefficient $\kappa \propto E_1E_2$ is too weak for generating biphotons, and the EIT slow light effect will always be dominant in previous systems, when $E_2$ is tuned in the weak coupling regime.

An EIT-based system can not only assist the SFWM nonlinear process but also enables the system to serve as a non-Hermitian platform. In order to realize flexible non-Hermitian control while simultaneously maintain stable biphotons generation, a dressing beam $E_3$ was applied to form a $V$-type EIT configuration. In such EIT-based non-Hermitian structure, the effective Hamiltonian of two coupled energy levels system $|2\rangle \to |4\rangle$ is shown as

$$H_{eff} = \begin{bmatrix} -i\Gamma_{21} & -\Omega_3/2 \\ -\Omega_3/2 & \Delta_3 - i\Gamma_{41} \end{bmatrix}, \qquad (1)$$

where $\Gamma_{ij} = (\Gamma_i + \Gamma_j)/2$ are the decoherence rate between energy levels $|i\rangle$ and $|j\rangle$; $\Gamma_i$ represents the decay rate of the $|i\rangle$. $\Omega_3$ is the Rabi frequency induced by the optical field $E_3$, and $\Delta_3$ is corresponding detuning. Eq. (1) is a typical non-Hermitian Hamiltonian containing coupling and dissipative terms, quantified by $\Omega_3$ and $\Gamma_{41/21}$ respectively, with the eigenvalues

$$\delta_\pm = -\Delta_3/2 + i\Gamma_{eff} \pm \sqrt{|\Omega_3|^2/4 + (\Delta_3/2 - i\Gamma_{diff})^2}. \qquad (2)$$

where $\Gamma_{eff} = (\Gamma_{41} + \Gamma_{21})/2$ is the effective decoherence rate defined as the average values of the holistic loss factor of system; $\Gamma_{diff} = (\Gamma_{41} - \Gamma_{21})/2$ is the difference between the two decoherence rates.

It is seen from Eq. (2) that the non-Hermitian control of the system is realized by varying $\Omega_3$, which quantifies the light-matter coupling strength in the transition $|2\rangle \to |4\rangle$. Meanwhile, the biphoton process also relies on suchtransition. As tuning $\Omega_3$ affects both the non-Hermitian features and the biphoton properties in the same manner, the optical response, i.e., Rabi oscillation and decay features, in the biphoton process can be actively controlled by the non-Hermitian properties, and simultaneously serves as a witness for the non-Hermitian properties in both eigenvalues and eigenfuctions. Fig. 1c illustrates the eigenenergy $\delta$ of $E_S$. Around EP (the two eigenvalues coalesce) as shown in Fig. 1d, $\delta_\pm$ can be tuned to dual coherent channels with strong and weak coupling by changing $\Omega_3$. Here, each biphoton coherent channel could generate a pair of frequency-entangled photons. The quantum interference of the multiple coherent channels corresponds to high-dimensional frequency-entangled entanglement[40,41], and the corresponding Hamiltonian is then extended as $H_I = i\hbar\kappa\Phi \sum_{j=1}^{n} c_j \hat{a}_{Sj}^\dagger \hat{a}_{ASj}^\dagger + H.C.$.

The experimental system provides an efficient platform to fine tune $\Omega_3$ and $\Delta_3$ in the non-Hermitian Hamiltonian Eq. (1) with the laser parameters of $E_3$. This constitutes a unique advantage over other light-matter coupling systems that lack tunability. Before discussing the non-Hermitian properties of the EP, we first introduce the three different regimes of the biphoton correlation classified by the interplay between nonlinear and linear susceptibilities in the SFWM optical responses. The diagram of $\chi_{AS}^{(3)}$ in weak coupling, EP and strong coupling regimes are shown in Fig. 2b, d, respectively, in which, the corresponding linear susceptibilities $\chi_{AS}$ have same profiles but are stronger with several order of magnitude. This is of high importance in this work, because only the nonlinear susceptibility $\chi_{AS}^{(3)}$ can be described by the non-degeneracies of the Hamiltonian Eq. (1) which contains the non-Hermiticity and the EP, while the linear one mainly results in varying complex refractive index with the EIT slow-light effect[39]. Therefore, for observing EP, such nonlinear susceptibility $\chi_{AS}^{(3)}$ need be strong enough compared to the linear one, otherwise the system characteristic would be predominantly determined by the linear response that prevents observing the non-Hermiticity around the EP. We plot the eigenvalues as a function of $\Omega_3/\Gamma_{41}$ at $\Delta_3 = 0$ in Fig. 3a, which is divided into three different regions named as R1, R2 and R3 herein. The regions of R1 and R3 is dominated by the nonlinear susceptibility and belongs to the so-called Rabi oscillation regime, in which the effective Rabi frequency $\Omega_e$ and effective decoherence rate $\Gamma_{eff}$ are smaller than the phase-matching bandwidth $\Delta\omega_g$, i.e., $\Delta\omega_g > \{|\Omega_e|, \Gamma_{eff}\}$, so the biphoton correlation is mainly determined by the $\chi_{AS}^{(3)}$ in dressed-state picture[36,39]; where the $\Delta\omega_g$ approximately calculate as $\Delta\omega_g \approx \pi|\Omega_3|^2/(OD\Gamma_{41})$. In this regime, the ground level $|2\rangle$ splits into two dressed states separated by $\hbar\Omega_e$, and the eigenstates of biphoton are manipulated to dual coherent channels.

In the case of region R1, considering a real effective Rabi frequency $\Omega_e$ (depending on the comparison between $\Omega_3$ and $\Gamma_{diff}$ in Eq. (2) defined as $\Omega_e = \delta_+ - \delta_-$), the Glauber correlation function of biphotons can be calculated as (details in the Supplemental Material)

$$G^{(2)}(\tau) = W_1 e^{-2\tau\Gamma_{eff}/W_D}[1 - \cos(\Omega_e\tau/W_D)]\Theta(\tau) \qquad (3)$$

where $W_1$ contains all the constants and slowly varying terms; $\tau$ is the relative time delay in the bi-photon state, defined by $\tau = t_{AS} - t_S$ with $t_{AS/S}$ representing the triggering time of each SPCM detector; Heaviside step function here is defined as $\Theta(\tau) = 1$ for $\tau \geq 0$, otherwise $\Theta(\tau) = 0$. The two-photon correlation exhibits damped Rabi oscillations, resulting of the destructive interference between two resonance channels of biphoton generation.

Meanwhile, in R3, the $G^{(2)}$ of biphoton state can be expressed as

$$G^{(2)}(\tau) = \frac{W_1}{|\Omega_e|^2} e^{-2\tau \Gamma_{eff}/W_D} \left[ e^{\Omega_e \tau/W_D} - e^{-\Omega_e \tau/W_D} \right]^2 \Theta(\tau) \qquad (4)$$

Since the coupling beam is very weak, two types of biphotons are generated with the same central frequency but are associated with different linewidths $\Gamma_e$ (defined as Im[$\delta$]), $\Gamma_{e+} = \Gamma_{eff} + \Omega_e/2$ and $\Gamma_{e-} = \Gamma_{eff} - \Omega_e/2$ due to $\Omega_3 < 2\Gamma_{diff}$. The existence of linewidth bifurcation and a common frequency is the typical character of the weak light-matter coupling regime. The interference between these two types of the biphotons waveform gives a manifested antibunching-exponential-decay (details in the Supplemental Material).

On the other hand, the region of R2 is reached when $\Delta\omega_g < \{|\Omega_e|, \Gamma_{eff}\}$, where the biphoton correlation is mainly determined by the phase-matching spectrum and the linear susceptibility play the important role. The biphoton is then in the so-called group delay regime featured by a single exponential decay (details in the Supplemental Material). The Region of R2 also contains both the physical processes of R1 and R3, the visibility of which is, however, affected by the linear slow-light effect. Although the longer group-delay time can partially wash the mechanism of double coherent channels of biphoton generation, leading to a deterioration of the observability of the Rabi oscillation as well as other features resulting from Eq. (1), these main features still exist and are observable, which robustly enables the demonstration of different light-coupling regimes and the EP, as will be evidenced later.

The temporal correlations of the biphoton state are observed via measuring the two photons coincidence counting, as shown in Fig. 3b, e corresponding to a series of different power of $E_3$, whereas the solid curves in Fig. 3b-d are theoretical simulations with the same experimental parameters. In Fig. 3b at the $E_3$ power of 15 mW with $\Delta_3 = 0$, the waveform profile of biphotons exhibits damped Rabi oscillations amongst the two detection events of a period near 4.5 ns, characterizing the strong coupling regime. Some background counts result from uncorrelated single photons from Raman scattering and dark noises. When the power of $E_3$ is tuned to 1 mW in Fig. 3c, the waveform profile of biphoton correlation shows a single exponential decay, which indicates that there is only one single coherent channel of two photon generation showing coalesced frequency and dissipation linewidth. This is thereby the EP. It is important to notice that the perfect single exponential decay only applies in the ideal situation presented by the blue theoretical curve in Fig. 3c, which displays $G^{(2)}(0) \neq 0$ and is therefore not antibunching. The experimental data, though still showing $G^{(2)}(0) = 0$ followed by a sharp rise due to the technical limit of our laser system (which can be improved by better stabilizing the laser wavelength), can nevertheless be well fitted by the single exponential theoretical curve. In this sense, we identify this type of waveform as non-antibunching, which is distinguished from the exponential decay features in R3 showing an actual antibunching even ideally. Indeed, the region of R3 is reached by decreasing the power of $E_3$ to 0.25 mW in Fig. 3d, where the waveform shows a slow climb from $\tau = 0$ resulting from the superposition of two eigenstates in the weak coupling regime. Note that, herein the terms "Rabi oscillation" in R1, "single exponential decay" for the EP in R2 and "antibunching exponential decay" in R3 are just names to classify different features rather than representing full properties of the temporal waveform. In addition, the corresponding energy levels diagrams of dual coherent channels $\text{Re}[\delta_\pm]$ and $\text{Im}[\delta_\pm] = \Gamma_{e\pm}$ are given in Fig. 3f, g.

Questions may rise from the fact that the linear susceptibility in the SFWM, which is more important than the nonlinear process for the region of R2, could also lead to a feature of single exponential decay in biphoton correlation function. We exclude this possibility by measuring the biphoton correlations of a series of points D, E and F approximately within R2, as shown in Fig. 3e. It is obvious that despite being in R2, the points D and E still show the Rabi oscillations demonstrating the strong coupling regime. The only qualitative difference with A is that the minimums of the oscillations do not reach zero due to the superimposition with the exponential decay features induced by the slow-light effect. Meanwhile, the point F still show the antibunching exponential decay features indicating the weak coupling regime. Therefore, the linear process in R2 does not completely wash out the non-Hermitian coupling features by the $\chi_{AS}^{(3)}$. Note that such non-Hermiticity inherently endowed within the light-matter coupling process would not be revealed in the previous systems, unless one enables the system highly controllable for both the atomic SFWM and the non-Hermitian control. Indeed, a traceable variation can be clearly identified following the A-D-B-E-C-F pathway, displaying a quasi-continuous transition from the strong to the weak coupling regimes as shown in Fig. 3a with experimental circles, in which the transition point B is unambiguously the EP. Here, the circles are obtained by fitting the corresponding eigenvalues of experimental waveforms of biphoton correlations based on the Eq. (3), (4).

The photon-atom interface allows flexible coherent control of the correlation waveform of biphotons around EP, which are important for practical applications of the entangled states. Fig. 4a shows the temporal correlations of biphoton measured by scanning $\Delta_3$ in the weak coupling regime. When varying the eigenvalue slightly away from the EP and along the axis normal to the plane of $\Delta_3 \neq 0$ in Fig. 1c, the real and imaginary parts of the eigenvalues do not simultenously coalese, as shown in Fig. 4b. The nonzero value of $\Delta_3$ induces two resonance channels of different frequencies, which coexist with the absorptive-channels of SFWM. In Fig. 4a, the starting point A is chosen at $\Delta_3 = 0$ and exhibits antibunching exponential decay. When increasing $\Delta_3$ from 0 to 100 MHz, the resulting curves of B-F show the oscillation features resulting from the nonzero $\Delta_3$ with dissipation. The capability of varying the $\Omega_3$ and $\Delta_3$ independently manifests the merits of the system which allows two-dimensional tunability around the EP.

The non-classical behavior of the biphoton correlation can be verified by the violation of Cauchy-Schwarz inequality (details in the Supplemental Material). We obtain the maximum factor $R_2$ of $116 \pm 13$ for violating the Cauchy-Schwarz inequality in Fig. 4b. Such non-Hermitian features can be expanded to *N*-order EP and *N*-dimensional entanglement with the multiple coherent channels induced by *N*-dressing effects with atomic coherence[40]. By adding the dressing effect, a multi-period Rabi oscillation of biphoton quantum correlation is illustrated in the Supplemental Material.

In conclusion, we demonstrated the transition from strong to weak light-matter coupling regimes in the biphoton process with atomic system, and observed unambiguously the non-Hermitian EP featured by the coalesced singlet state. The merit of our results lies in the proper demonstration of non-Hermiticity and EP of a naturally existing, unstructured light-matter coupling system, especially in nonlinear and entanglement generation processes. These results provide a unique method to efficiently generate quantum entanglement and to control quantum optical processes based on the most comprehensive interpretation of light-matter interaction as a natural non-Hermitian system. Such control mechanism allowed by quantum optical manipulation would lead to profound discoveries in the area of quantum non-Hermitian physics and phenomena. It is foreseeable that the application of non-Hermitian physics beyond the artificially designed photonic structures, initiated by this work, will be extended to various natural systems to reach novel platforms for the control of physical states and quantum information.

## Methods
### Experimental implementation

The generation of biphoton employs an $^{85}$Rb vapor cell of length 7 cm with a temperature of 60°C as sketched in Fig. 2a. In the presence of three coaxial cw laser beams, including one pump ($E_1, \omega_1, \vec{k}_1$), one coupling ($E_2, \omega_2, \vec{k}_2$) and one extra dressing ($E_3, \omega_3, \vec{k}_3$), backward photon pairs $E_S$ ($\omega_S, \vec{k}_S$) and $E_{AS}$ ($\omega_{AS}, \vec{k}_{AS}$) are emitted with an intersection angle about 4° to the axial direction of the input lasers by phase matching condition, $\Delta\vec{k} = (\vec{k}_S + \vec{k}_{AS}) - (\vec{k}_1 + \vec{k}_2) = 0$. Employing the external cavity diode lasers, $E_1$ (795 nm, vertical polarization) puts a large detuning $\Delta_1$ (2 GHz), which results in the quantum atomic noise being largely suppressed and the atomic population resides primarily at the ground state. $E_2$ (780 nm, horizontal polarization) and $E_3$ (780 nm, vertical polarization) with near-resonant conditions drives the transition $|2\rangle \to |4\rangle$ simultaneously but in counter-propagating directions.

Due to thermal motion of atoms, the eigenenergies of biphotons could be broaden by applying the Doppler-broadening effect in optical responses. The Doppler width calculates near 539 MHz, and the atomic density is $2.5\times10^{11}$ cm$^{-3}$. To avoid unwanted accidental trigger events, single-band filters and narrowband etalon Fabry–Perot cavities are employed to filtrate the uncorrelated single photons. The bandwidth, transmission efficiency, and extinction ratio of the filters are nearly 350 MHz, 80%, and 60 dB, respectively. The bandwidth of the Fabry-Perot cavity is near 500 MHz. With a temporal bin width of 0.0244 ns, electronic pulses from SPCMs of photons are inputting into the time-to-digit converter, where the maximum resolution time of our recording card is 813 fs. The coupling efficiency of the fibers and the detection efficiency of the SPCMs is 70% and 40%, respectively. Besides, an additional optical-pumping beam $E_{op}$ with the resonance transition $|2\rangle \to |3\rangle$ is added to decrease the noise from the Raman scattering[36,38]. Due to the large detuning of photon $E_S$, the group velocity of $E_S$ approaches $c$ while $E_{AS}$ is generated with EIT slow-light effect. In this case, the detected $E_S$ and $E_{AS}$ serve as the start triggering photon and stop triggering photon, respectively. Therefore, the two-photon coincidence count is recorded, so that the temporal correlation of biphoton is obtained.

**Acknowledgements**
This work was supported by the National Key Research and Development Program of China (2017YFA0303700, 2018YFA0307500), National Natural Science Foundation of China (12174302, 61975159, 61605154, 11604256, 11804267, 11904279, 12074303), Shaanxi Key Science and Technology Innovation Team Project (2021TD-56).

**Conflict of interest**
The authors declare no competing financial interests.

**Contributions**
Y.C., F.L and Y.Z. conceived the idea and supervised the project. K.L., supervised by Y.C. and Y.Z., performed the experiment, theoretical derivations, and numerical calculations with the help from J.Y., Z.F. and F.L., K.L., Y.C. and F.L. wrote the manuscript with contributions from all other authors. All contributed to the discussion of the project and analysis of experimental data.

32. Li, Y. Q. & Xiao, M. Observation of quantum interference between dressed states in an electromagnetically induced transparency. *Phys. Rev. A* **51**, 4959 (1995).
33. Wu, B. et al. Slow light on a chip via atomic quantum state control. *Nat. Photon.* **4**, 776–779 (2010).
34. Pang, M. et al. All-optical bit storage in a fibre laser by optomechanically bound states of solitons. *Nat. Photon.* **10**, 454–458 (2016).
35. Yan, H. et al. Generation of narrow-band hyperentangled nondegenerate paired photons. *Phys. Rev. Lett.* **106**, 033601 (2011).
36. Li, K. et al. Dressing-Shaped Rabi Oscillation of Photon Pairs in Hot Atomic Ensemble. *Adv. Quantum Technol.* **3**, 2000098 (2020).
37. Mei, Y. et al. Einstein-podolsky-rosen energy-time entanglement of narrow-band biphotons. *Phys. Rev. Lett.* **124**, 010509 (2020).
38. Shu, C. et al. Subnatural-linewidth biphotons from a Doppler-broadened hot atomic vapour cell. *Nat. Commun.* **7**, 12783 (2016).
39. Du, S. Wen, J. & Rubin, M. H. Narrowband biphoton generation near atomic resonance. *JOSA B* 25, C98–C108 (2008).
40. Liu, Y. et al. Atomic-coherence-assisted multipartite entanglement generation with dressing-energy-level-cascaded four-wave mixing. *Phys. Rev. A* **106**, 043709 (2022).
41. Zhang, D. et al. Generation of high-dimensional energy-time-entangled photon pairs. *Phys. Rev. A.* **96**, 053849 (2017).
**List of Figures**

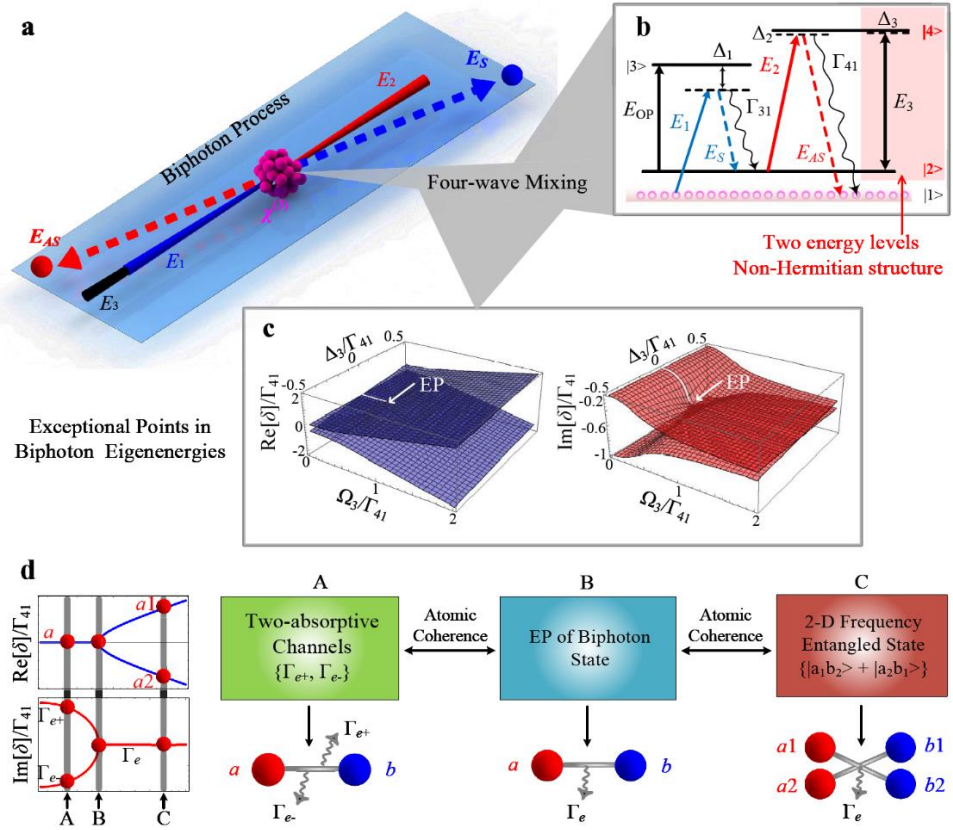

**Fig. 1 The biphoton as a non-Hermitian system. a** Schematic of entangled biphotons generation via SFWM process. **b** Two-Λ EIT-based energy levels sketch of the SFWM. **c** The real and imaginary parts of eigenenergy nature $\delta$ of biphotons. $\Gamma_{21} = 0.2\Gamma_{41}$. Two eigenstates are merged into one at EP with $\Omega_3 = 0.8\Gamma_{41}$ and $\Delta_3 = 0$. **d** Around EP (position B), the eigenstates of biphoton can be manipulated to multi-channel properties (positions A and C) by atomic coherence.

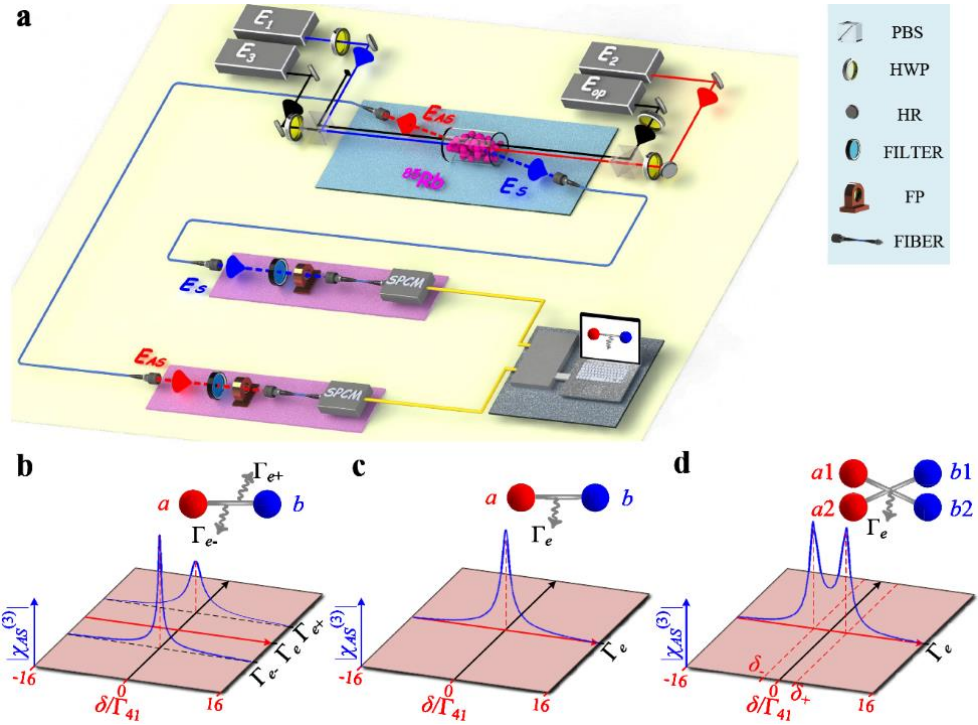

**Fig. 2 The principle of the experiment. a** Experiment setup. Using optical lenses, three coaxial driving fields $E_1$, $E_2$ and $E_3$ are coupled into the center of rubidium atomic vapor cells; PBS, polarization beam splitter; M, mirror; $\lambda/2$, half-wave plate; the spatially separated biphoton $E_S$ and $E_{AS}$ are coupled into the data acquisition system by the single-mode fibers. Through the single-band filters and narrowband etalon Fabry–Perot cavity (FP), the generated photons are detected by the synchronized single-photon counting modules (SPCM); All detected events are measured by a fast-time acquisition card with a computer. **b-d** Diagram of the nonlinear susceptibility in different coupling strength.

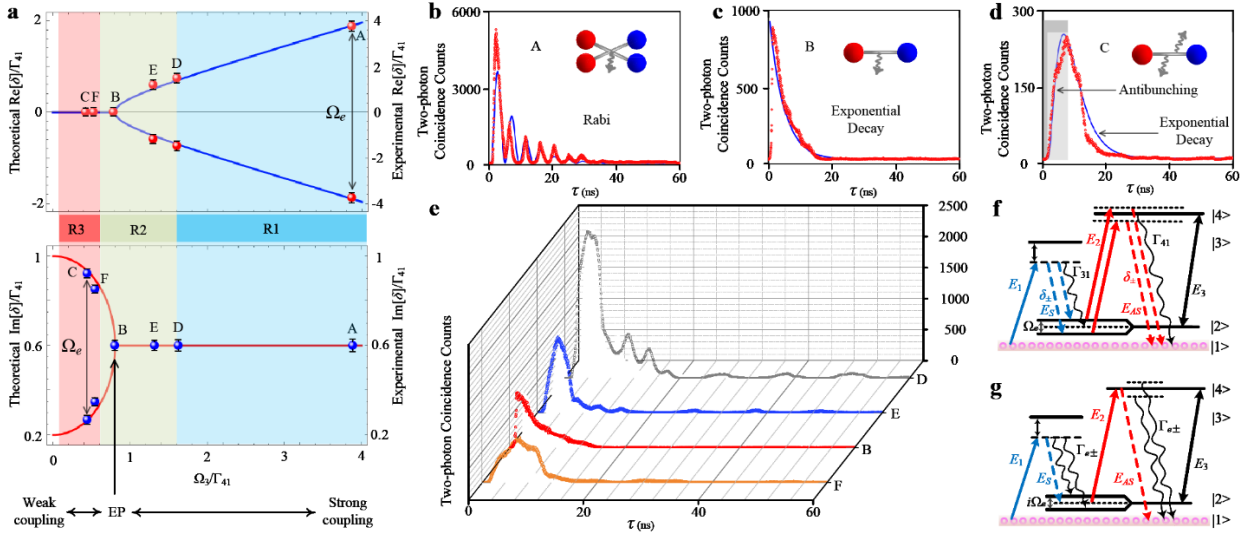

**Fig. 3 Demonstration of the EP. a** Calculated real and imaginary parts of $\delta$ with $\Delta_3 = 0$ by scanning $\Omega_3$. R1-R3, three different property regions Rabi oscillation, group delay and antibunching-exponential-decay, respectively. The solid circles of different colors labeled by letters A-F correspond to the situations labeled by the same letters in **b**-**e**. **b** Two-photon coincidence counting measurements. Collected about 10 min and each bin corresponds to a 0.2 ns time interval. The power of $E_1$, $E_2$, and $E_3$ are P1 = 4 mW, P2 = 3 mW, and P3 = 15 mW, respectively. $\Delta_1 = 2$ GHz. $\Delta_2 = 200$ MHz. $\Delta_3 = 0$. OD = 6.8. **c** P3 = 1 mW; (d) P3 = 0.25 mW. Two arrows are drawn in **d** to identify the antibunching and exponential decay feature in the graphs. **e** Biphoton temporal correlation around EP by scanning $\Omega_3$. **f** and **g** Energy levels sketch of the SFWM dual-coherent channels.

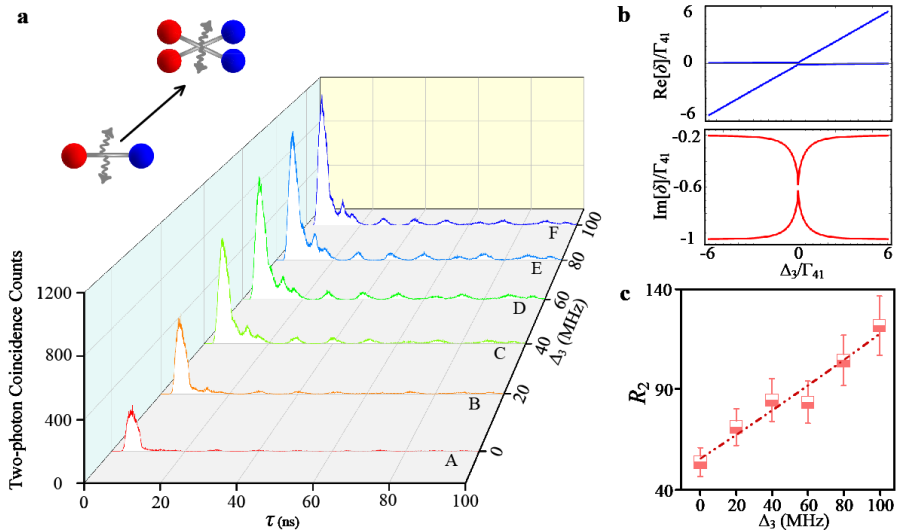

**Fig. 4 Manipulation of biphoton correlation. a** Two-photon coincidence counting measurements by scanning $\Delta_3$ from 0 to 100 MHz, when $E_3$ set as 0.6 mW. **b** The real and imaginary parts of $\delta$ with $\Omega_3 = 0.7\Gamma_{41}$ by scanning $\Delta_3$. **c** The factors for violating the Cauchy-Schwarz inequality.

# Supplementary Information for "Direct Manipulation of quantum entanglement from the non-Hermitian nature of light-matter interaction"

Kangkang Li[1,2], Yin Cai[1, ‡], Jin Yan[1], Zhou Feng[1], Fu Liu[1], Lei Zhang[1], Feng Li[1, †], Yanpeng Zhang[1, *]

[1]*Key Laboratory for Physical Electronics and Devices of the Ministry of Education & Shaanxi Key Lab of Information Photonic Technique, Xi'an Jiaotong University, Xi'an 710049, China*
[2]*State Key Laboratory for Artificial Microstructure and Mesoscopic Physics and Frontiers Science Center for Nano-optoelectronics, School of Physics, Peking University, 100871 Beijing, China*

[*]Corresponding author: caiyin@xjtu.edu.cn; felix831204@xjtu.edu.cn; ypzhang@mail.xjtu.edu.cn.

**Supplementary Note 1**

**The third-order nonlinear susceptibility of spontaneous four-wave mixing processes**

In the EIT-based spontaneous four-wave mixing (SFWM) process, the physics of multi-channel biphotons generation can be explained by the nonlinear susceptibility $\chi^{(3)}$ (e.g., Ref. [S1]). The Hamiltonian of SFWM can be written as: $H_I = -(e\hat{x})\vec{e}_x \cdot \vec{E} = -\frac{\hbar}{2}([(\Delta_1 - \Delta_S)|2\rangle\langle 2| + \Delta_1|3\rangle\langle 3| + (\Delta_1 - \Delta_S - \Delta_2)|4\rangle\langle 4|] + \hbar[\Omega_1|1\rangle\langle 3| + \Omega_S|3\rangle\langle 2| + \Omega_2|2\rangle\langle 4| + \Omega_{AS}|4\rangle\langle 1| + \Omega_3|4\rangle\langle 2|])$. According to motion equation, the density–matrix equations can be obtained to be $\frac{\partial}{\partial t}\rho = -\frac{i}{\hbar}[H_I, \rho] - \Gamma\rho$:

$\frac{\partial \rho_{11}^{(r)}}{\partial t} = -\Gamma_{11}\rho_{11}^{(r)} + i\left[\Omega_1^* \rho_{31}^{(r-1)} - \Omega_{AS}\rho_{41}^{(r-1)}\right]/2;$

$\frac{\partial \rho_{31}^{(r)}}{\partial t} = -(i\Delta_1 + \Gamma_{31})\rho_{31}^{(r)} + i\left[\Omega_1\left(\rho_{11}^{(r-1)} - \rho_{33}^{(r-1)}\right) + \Omega_S \rho_{21}^{(r-1)}\right]/2;$

$\frac{\partial \rho_{21}^{(r)}}{\partial t} = -(i\Delta_1 - i\Delta_S + \Gamma_{21})\rho_{21}^{(r)} + i\left[(\Omega_2^* + \Omega_3^*)\left(\rho_{41}^{(r-1)} - \rho_{22}^{(r-1)}\right) + \Omega_S^* \rho_{31}^{(r-1)}\right]/2;$

$\frac{\partial \rho_{41}^{(r)}}{\partial t} = -(i\Delta_1 - i\Delta_S + i\Delta_2 + \Gamma_{21})\rho_{41}^{(r)} + i\left[(\Omega_2^* + \Omega_3^*)\left(\rho_{21}^{(r-1)} - \rho_{44}^{(r-1)}\right) + \Omega_{AS}\rho_{11}^{(r-1)}\right]/2.$

We applied the perturbation theory to obtain the density-matrix elements under the weak-field and steady state ($\partial\rho/\partial t \approx 0$) approximations, where the density matrix elements $\rho^{(0)}, \ldots, \rho^{(3)}$ are obtained step by step. The function of $\chi^{(3)}$ in SFWM is from the perturbation chain

$$\rho_{11}^{(0)} \xrightarrow{\omega_1} \rho_{31}^{(1)} \xrightarrow{\omega_S} \rho_{21}^{(2)} \xrightarrow{\omega_2} \rho_{41}^{(3)}, \text{(S1.1)}$$

where $\chi^{(3)} = N\mu_{14}\rho_{41}^{(3)}/(\varepsilon_0 E_1 E_S^* E_2)$. In hot vapor cells, the atoms move randomly with a Maxwell-Boltzmann velocity distribution. This thermal motion causes both Doppler shift and broadening effects in atom-light interaction. In the EIT configurations, the Doppler shift can be neglected due to the same Doppler direction (e.g., Ref. [S1]). However, the eigenenergies of biphotons could be broaden in optical responses. Therefore, by applying the Doppler-broadening coefficients, we normalized the third-order nonlinear susceptibility in the dressed-state picture of the hot atomic systems as

$$\chi_{AS}^{(3)} = \int f(v) \frac{N\mu_{13}\mu_{23}\mu_{24}\mu_{14}}{\varepsilon_0 \hbar^3 (\Gamma_{31}+i\Delta_{1D})d_{EIT}(\Gamma_{41}+i\Delta_{2D}+iW_D\delta+d_2)} dv. \quad (S1.2)$$

where $\Gamma_{ij} = (\Gamma_i + \Gamma_j)/2$ is the decoherence rate between $|i\rangle$ and $|j\rangle$; $\mu_{ij}$ are the electric dipole matrix elements; $N$ is atomic density; $\varepsilon_0$ is the vacuum permittivity; $\hbar$ is Planck constant divided by $2\pi$; $d_2$ for dressing effect of $E_2$ can be viewed as a constant due to the stable power of $E_2$; $W_D$ is Doppler-broadening coefficient, that $W_D = 1 + v/c$; $v$ is the speed of atomic motion; $\Delta_{1D}$ and $\Delta_{2D}$ are the detuning of $E_1$ and $E_2$ that are defined as $\Delta_{1D} = \omega_{31} - W_D\omega_1$ and $\Delta_{2D} = \omega_{41} - W_D\omega_2$; $f(v)$ is the 1D Maxwell-Boltzmann distribution, that $f(v) = \sqrt{m_{Rb}v^2/2\pi k_B T} \times exp[-m_{Rb}/2k_B T]$; here $T$ is the cell temperature and $k_B$ is the Boltzmann constant (e.g., Refs. [S1] and [S2]). The spectral properties of biphoton is essential characteristic of optical responses, in which the natural spectral of generated biphoton ($\omega_S$, $\omega_{AS}$) can express as ($\varpi_S - \delta$, $\varpi_{AS} + \delta$) with ($\varpi_S$, $\varpi_{AS}$) being the central frequencies of biphoton and $\delta$ represent the eigenenergy window around the central frequencies. In Eq. (S1.2), the EIT-based non-Hermitian term $d_{EIT}$ of $\chi_{AS}^{(3)}$ can express as

$$d_{EIT} = \Gamma_{21} + iW_D\delta + \frac{|\Omega_3|^2}{\Gamma_{41}+iW_D\delta+i\Delta_{3D}}. \quad (S1.3)$$

where $\Delta_{3D}$ is the detuning of $E_3$ defined as $\Delta_{3D} = \omega_{41} - W_D\omega_3$; $\Omega_3$ is Rabi frequency of $E_3$. The corresponding eigenvalues in Eq. (S1.3) are calculated in Eq. (2). Assuming effective Rabi frequency $\Omega_e = \delta_+ - \delta_-$, an EP occurs when $\Omega_e$ is zero, as the two eigenvalues coalesce. We first consider the resonance situation of $\Delta_3 = 0$, as described by Fig. 1d, which is the middle cross-section of Fig. 1c. When $\Omega_3 > 2\Gamma_{diff}$, the strong coupling regime where $\delta_\pm$ display two eigenenergy $\delta_+ = i\Gamma_{eff} + \Omega_e/2$ and $\delta_- = i\Gamma_{eff} - \Omega_e/2$. Based on the energy conservation ($\omega_1 + \omega_2 = \omega_S + \omega_{AS}$) of biphoton, the SFWM is therefore has two resonance channels Re[$\delta_\pm$] with same effective decoherence rate $\Gamma_{eff}$. When $\Omega_3 < 2\Gamma_{diff}$, the system is in the weak coupling regime where the photon $E_S$ has two eigenenergy $\delta_\pm = i\Gamma_{eff} \pm i\Omega_e/2$. In this case, the SFWM get same resonance frequency window with two absorptive channels Im[$\delta_\pm$]. Therefore, the corresponding effective resonance linewidth can rewritten as $\Gamma_{e\pm} = \Gamma_{eff} \pm \Omega_e/2$, which can explain as the effective resonance linewidth $\Gamma_{eff}$ are split by the imaginary effective Rabi frequency $\Omega_e$.

Therefore, the nonlinear susceptibility $\chi_{AS}^{(3)}$ with double-resonance SFWM channels, EP, and double-absorptive SFWM channels, are shown in Supplementary Fig. 1a, c, respectively. In the double-resonance SFWM channels case, the generated biphoton state can be described as the two-dimensional frequency entangled state as

$$|\Psi\rangle = \left[N_1|\Psi_{S_+,AS_-}\rangle - N_2|\Psi_{S_-,AS_+}\rangle\right]. \quad (S1.4)$$

where $N_1$ and $N_2$ satisfy $N_1^2 + N_2^2 = 1$ and $N_1^2 - N_2^2 = 0$ because of the normalization and the destructive interference. Here, two types of biphotons are generated with the same linewidth $\Gamma_{eff}$ but are associated with different central frequencies: $\{\varpi_S + \Omega_e/2, \varpi_{AS} - \Omega_e/2\}$ and $\{\varpi_S - \Omega_e/2, \varpi_{AS} + \Omega_e/2\}$ as seen in Supplementary Fig. 1a. In EP as shown in Supplementary Fig. 1b, the eigenenergy of biphotons coalesce at the same linewidth and central frequency. In double-absorptive channels case at Supplementary Fig. 1c, the eigenvalues of $\chi_{AS}^{(3)}$ are obtained with the same central frequency but are associated with different linewidths: $\Gamma_{eff} + \Omega_e/2$ and $\Gamma_{eff} - \Omega_e/2$. For showing the imaginary properties, we changed the continuous-frequency-mode $\delta$ to imaginary basis with $\delta_{\text{Im}} = i\delta$ as shown in Supplementary Fig. 1d-f. In this case, the $\chi_{AS}^{(3)}$ shown a clearly two linewidths in Supplementary Fig. 1f.

**Supplementary Note 2**

**The linear susceptibility of spontaneous four-wave mixing processes**

The natural spectral width of biphoton is determined by the linear optical response. In our system, $E_1$ is weak powered and puts a large detuning by $\Delta_1 = 2$ GHz. Therefore, by considering $\{|\Omega_2|^2, |\Omega_3|^2\} \ll \Delta_1$, the linear susceptibility of $E_S$ can be approximated by $\chi_{AS}$, which means the group velocity of $E_S$ approaches $c$. Consequently, the linear susceptibility of $E_{AS}$ can be written as

$$\chi_{AS}^{(1)} = \int f(v) \frac{-N\mu_{14}^2(W_D\delta - i\Gamma_{21} - i\Delta_{3D})}{\varepsilon_0 \hbar[(W_D\delta - i\Gamma_{41})(W_D\delta - i\Gamma_{21} - i\Delta_{3D}) - |\Omega_3|^2 - c_2]} dv. \quad (S2.1)$$

Using the slow-light effect, the group velocity of $E_{AS}$ can approximately write as follows

$$v_g = \frac{2k_{14}Lc(|\Omega_3|^2 - \Gamma_{41}\Gamma_{21})^2}{\omega_{14} OD \Gamma_{41} W_D [|\Omega_3|^2 + c_2 - \Gamma_{21}^2]}, \quad (S2.2)$$

where it can be defined by the function of $c/[n + \delta(dn/d\delta)]$; $n = \sqrt{1 + Re[\chi_{AS}]}$ is the refractive index; $k_{14}$ represents the complex wavenumber of the resonant transition; $c_2 \propto |\Omega_2|$ can be viewed as a constant due to the stable power of $E_2$; OD is optical depth and defined as $OD = N\sigma_{14}L$ with $\sigma_{14}$ of the on-resonance absorption cross section in the transition $|1\rangle \to |4\rangle$ that $\sigma_{14} = 2\pi\mu_{14}^2/(\varepsilon_0 \hbar \lambda_{14} \Gamma_{41})$; $L$ is the length of the atomic medium. The group velocity controls the transmission spectrum and dispersion profile of the generated photons. The complex wavenumber of $E_{AS}$ is written as $k_{AS} = k_{AS0} + \delta/v_g + i\alpha$, whereas, the real part is Raman gain and the imaginary part is EIT loss; $\alpha = 2N\sigma_{14}\Gamma_{41}\Gamma_{21}/(|\Omega_3|^2 + 4\Gamma_{41}\Gamma_{21})$; $k_{AS0}$ is the central wave-number of $E_{AS}$. Therefore, the phase-matching of biphoton generation is written as $\Delta k(\delta) = k_S - k_{AS} - k_1 + k_2 \approx \delta/v_g + i\alpha$. The correspond-ing phase-matched bandwidth determined by the sinc function, $\Delta\omega_g = 2\pi v_g/L \approx \pi|\Omega_3|^2/(OD\Gamma_{41})$, which means the $\Delta\omega_g$ can be effectively manipulated by the values of coupling strength $\Omega_3$ and optical depth OD.

## Supplementary Note 3
## Two-photon Glauber correlation function

The effective interaction Hamiltonian of the biphoton processing is defined as $H_I = i\hbar\kappa\Phi\hat{a}_S^\dagger\hat{a}_{AS}^\dagger + H.C.$. According to the first-order perturbation in the interaction picture, we know that the photon state at the output surface is approximately a linear superposition of $|0\rangle$ and $|\Psi\rangle$. We can only consider the two-photon part since vacuum $|0\rangle$ is not detectable. Combined with the quantized fields of generated photons and the frequency entanglement of the biphoton state, the biphoton amplitude (or wave function) in the time domain is given as (e.g., Ref. [S3])

$$\psi(\tau) = \frac{L}{2\pi}\int d\delta\, \kappa(\delta)\Phi(\delta)e^{-i\delta\tau}, \quad (S3.1)$$

where the nonlinear coupling coefficient $\kappa$ is defined as $\kappa(\delta) = -i\left(\sqrt{\varpi_S\varpi_{AS}}/2c\right)\chi_{AS}^{(3)}(\delta)E_1E_2$; the longitudinal detun-ing function $\Phi$ is expressed as $\Phi(\delta) = \text{sinc}(\Delta k(\delta)L/2)e^{-i\Delta k(\delta)L/2}$. The joint-detection probability of biphotons is calculated by Glauber's theory (e.g., Ref. [S3])

$$G^{(2)}(\tau) = |\langle 0|\hat{a}_{AS}\hat{a}_S|\Psi\rangle|^2 = |\Psi(\tau)|^2, \quad (S3.2)$$

where $\Psi(\tau) = \psi(\tau)e^{-i(\omega_1+\omega_2)t_S}$. From Eq. (S3.1) and Eq. (S3.2), we see that the convolution of $\kappa$ and $\Phi$ decides the biphoton waveform, thereby manifesting the different regions in Glauber second-order temporal correlations. Ignoring the deviation of the detection, the average biphoton coincidence counting rate is determined by

$$Rcc(\tau) = G^{(2)}(\tau) + R_S R_{AS}. \quad (S3.3)$$

where $R_S R_{AS}$ represent the background from uncorrelated single photons.

In the Rabi oscillation and dissipative oscillation regimes, the biphoton correlation is mainly determined by the nonlinear optical response $\chi^{(3)}$ so that we can treat the longitudinal detuning function $\Phi(\delta) \approx 1$ (e.g., Ref. [S3]). Furthermore, in the group delay regime, the biphoton wave function can be approximated as $\psi(\tau) \approx \frac{L}{2\pi}\int d\delta\, \kappa_0\Phi(\delta)e^{-i\delta\tau}$ where $\kappa_0$ is the nonlinear coupling constant. Therefore, when the EIT loss is significant, the biphoton temporal correlations follow an exponential decay as

$$G^{(2)}(\tau) = W_1 \kappa_0^2 v_g^2 e^{-2\alpha v_g \tau / W_D} \Theta(\tau). \quad (S3.4)$$

In dissipative oscillation regime, the temporal correlation of biphotons performence as antibunching exponential decay as shown in Fig. S2. In R3, two types of biphotons are generated with the same central frequency but are associated with different linewidths, $\Gamma_{eff} + \Omega_e/2$ and $\Gamma_{eff} - \Omega_e/2$. Thus, we first plotted the biphoton temporal correlations $G^{(2)}$ in these two different absorptive channels solely in Supplementary Fig. 2a, b, respectively. In this case, the waveform profiles of biphoton correlations show as a single exponential decay with only one single coherent channel, which are similar with the biphoton temporal correlation in Fig. 3c but have different decay rates. Via measuring the two-photon coincidence counting, these two types of biphotons are detected simultaneously. Here, two absorptive channels of biphotons exist quantum interference effect, which results the waveform profile of $G^{(2)}$ performs as antibunching exponential decay in Supplementary Fig. 2c. For verifying the interference between two absorptive channels of biphotons, we shown the biphoton correlation in imaginary space by changing $\delta$ to imaginary basis with $\delta_{\text{Im}} = i\delta$ as shown in Supplementary Fig. 2d. In this case, a clearly periodic oscillation among two linewidths $\Gamma_{eff} + \Omega_e/2$ and $\Gamma_{eff} - \Omega_e/2$ are observed.

**Supplementary Note 4**

**Biphoton Cauchy-Schwarz inequality and high-dimensional entanglement**

To characterize the nonclassical property of the photon pairs source, we confirm its violation of the Cauchy-Schwarz inequality. The calculated factor of Cauchy-Schwarz inequality can be expressed as

$$R_2 = \frac{\left[g^{(2)}_{S,AS}(\tau)\right]^2}{\left[g^{(2)}_{S,S}(0)\right]\left[g^{(2)}_{AS,AS}(0)\right]} \leq 1. \quad (S4.1)$$

The normalized cross-correlation function $\left[g^{(2)}_{S,AS}(\tau)\right]$ is obtained by normalizing the two-photon coincidence counts to the flat background. The auto-correlations of the photon $E_S$ is measured using a fiber beam splitter with $\left[g^{(2)}_{S,S}(0)\right] = 1.6 \pm 0.2$. The autocorrelation of photons $E_{AS}$ have maximum values of 2. In our experiments, the background nonzero floor in wave packets are a result of accidental coincidence between uncorrelated single photons.

Several experimental platforms in photonics have been considered for the creation of highly entangled states, i.e., energy time, time bins, amplitude phase, orbital angular momentum, frequency modes and so on. Based on non-Hermitian structure in our manuscript, we manipulated the time-energy entangled biphoton to multiple coherent channels with strong and weak coupling. Here, each biphoton coherent channel could generate a pair of frequency-entangled photons. Therefore, the multiple coherent channels of biphoton process result the multi-frequency modes feature of biphoton states, which corresponds to multiple Herbert spaces. The quantum interference of the multiple coherent channels induced by non-degeneracies around EPs corresponds to high-dimensional entanglement.

Without changing the experimental configuration, the coupling field $E_2$ in Fig. 1b is tuned to strong coupling by increasing its laser power, in which, the generated dressed-state $|\Psi_{S_+,AS_-}\rangle$ could be split a second time as $[|\Psi_{S_{++},AS_{--}}\rangle + |\Psi_{+-,AS_{-+}}\rangle]/\sqrt{2}$ and the EIT-based non-Hermitian term $d_{\text{EIT}}$ of $\chi^{(3)}_{AS}$ with the double dressing effect can be rewritten as follows

$$d_{\text{EIT}} = \left[ \begin{array}{c} \left( W_D \delta - \frac{\Delta_{3D}}{2} + \Omega_e + i\Gamma_e + \frac{|\Omega_2|^2}{i\Gamma_{41} + W_D \delta - \frac{\Delta_{3D}}{2} + \Delta_{2D}} \right) \\ \left( W_D \delta - \frac{\Delta_{3D}}{2} - \Omega_e + i\Gamma_e \right) \end{array} \right]. \quad \text{(S4.2)}$$

Similarly, the biphoton state with triple coherent channels in Supplementary Fig. 3a can be rewritten as a three-dimensional frequency-entangled state as

$$|\Psi\rangle = \left[ N_1 |\Psi_{S_{++},AS_{--}}\rangle - N_2 |\Psi_{S_{+-},AS_{-+}}\rangle - N_3 |\Psi_{S_{-},AS_{+}}\rangle \right], \quad \text{(S4.3)}$$

In this case, a multi-period Rabi oscillation of biphoton quantum correlation is observed as shown in Supplementary Fig. 3b. Therefore, in our experiments, we manipulated such high-dimensional frequency entangled biphoton state by shaping the waveform profiles of its temporal correlations, i.e., the number and length of oscillation periods. In Supplementary Fig. 3c, we evaluate the evolution of the real and imaginary parts of the eigenvalues in the parameter space $(\Omega_3, \Omega_2)$, setting $\Gamma_{21} = 0.2 * \Gamma_{41}$. The multiple coherent channels induced by double dressing effect around EPs are obtained, which also indicates a high-order EP of biphoton state.

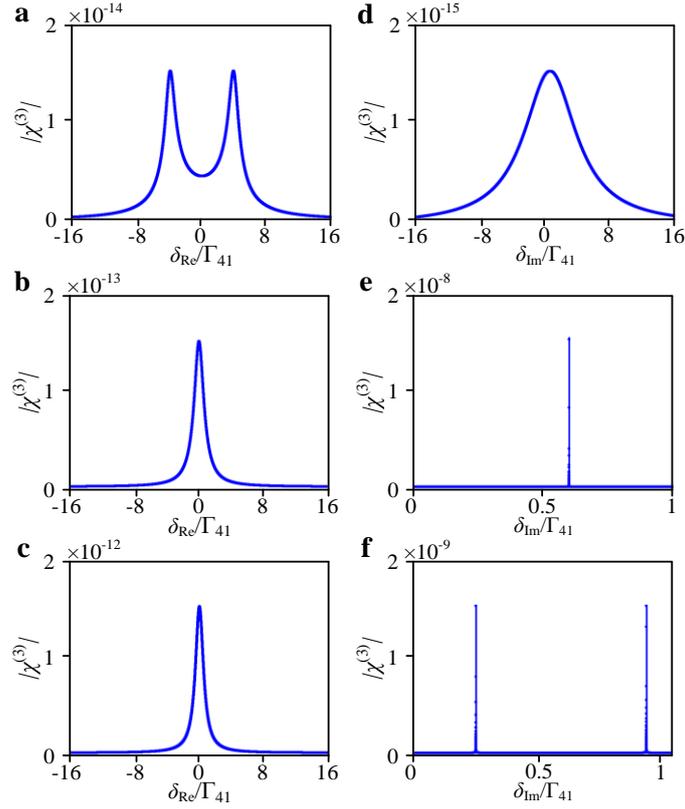

**Supplementary Figure 1. The nonlinear susceptibility. a-c** Diagram of the nonlinear susceptibility $\chi^{(3)}_{AS}$ with double-resonance SFWM channels, EP, double-absorptive SFWM channels, respectively. $\Gamma_{41} = \Gamma_{31} = 2\pi \times 6\text{MHz}$. $\Gamma_{21} = 0.2 \times \Gamma_{41}$. $\Gamma_{11} = 0.4 \times \Gamma_{41}$. $\Delta_1 = 52 \times \Gamma_{41}$. $\Delta_2 = 0$. $\Delta_3 = 0$. **d-f** same with **a-c**, but change the continuous-frequency-mode $\delta$ to imaginary basis with $\delta_{\text{Im}}$. **a, d** $\Omega_3 = 10 \times \Gamma_{41}$. **b, e** $\Omega_3 = 0.8 \times \Gamma_{41}$. **c, f** $\Omega_3 = 0.4 \times \Gamma_{41}$.

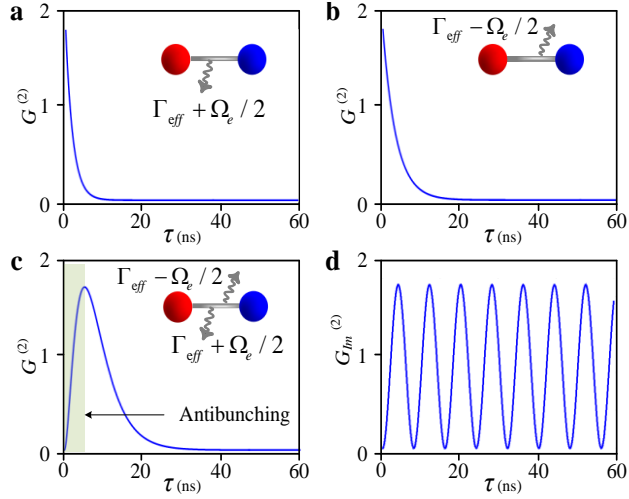

**Supplementary Figure 2. Theoretical simulation pictures of biphoton temporal correlations $G^{(2)}$. a** $G^{(2)}$ is plotted with single coherent channel, where the corresponding resonance linewidth is $\Gamma_{e+} = \Gamma_{eff} + \Omega_e/2$. $\Gamma_{41} = \Gamma_{31} = 2\pi \times 6\text{MHz}$. $\Gamma_{21} = 0.2 \times \Gamma_{41}$. $\Gamma_{11} = 0.4 \times \Gamma_{41}$. $\Delta_1 = 52 \times \Gamma_{41}$. $\Delta_2 = 0$. $\Delta_3 = 0$. $\Omega_3 = 0.4 \times \Gamma_{41}$. **b** $G^{(2)}$ is plotted with single coherent channel, where the corresponding resonance linewidth is $\Gamma_{e-} = \Gamma_{eff} - \Omega_e/2$. **c** $G^{(2)}$ is plotted with the quantum interference between two absorptive channels $\Gamma_{e\pm}$. **d** same with **c**, but change the $\delta$ to imaginary basis with $\delta_{\text{Im}}$. The biphoton temporal correlations in imaginary space $G_{Im}^{(2)}$ is plotted with the quantum interference between two absorptive channels.

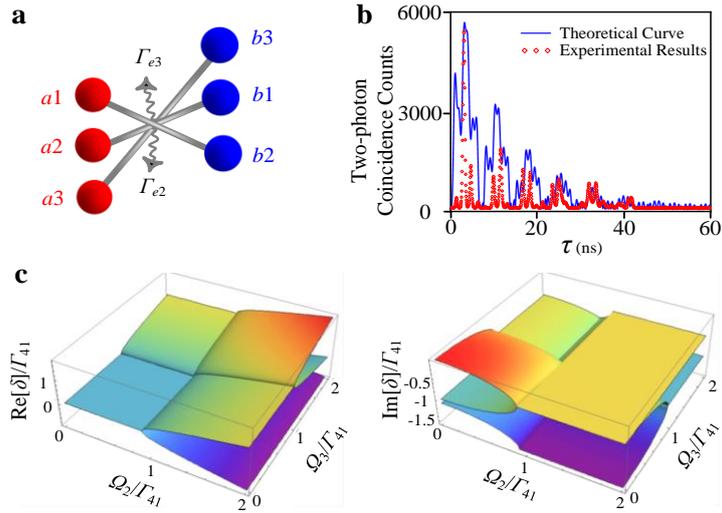

**Supplementary Figure 3. The observation of high-dimensional entanglement. a** Three-dimensional energy-time-entangled biphoton state. **b** Two-photon coincidence counting measurements. Collected about 10 min and each bin corresponds to a 0.2 ns time interval. The power of $E_1$, $E_2$, and $E_3$ are P1 = 4 mW, P2 = 10 mW, and P3 = 15 mW, respectively. $\Delta_1 = 2$ GHz. $\Delta_2 = 100$ MHz. $\Delta_3 = 0$. **c** The real and imaginary parts of $\delta$ by scanning $\Omega_3$ and $\Omega_3$.

## Supplementary References